%
%
\input lanlmac.tex
\overfullrule=0pt
\input mssymb.tex
\writedefs
%
\input epsf
\def\fig#1#2#3{%
\xdef#1{\the\figno}%
\writedef{#1\leftbracket \the\figno}%
\nobreak%
\par\begingroup\parindent=0pt\leftskip=1cm\rightskip=1cm\parindent=0pt%
\baselineskip=11pt%
\midinsert%
\centerline{#3}%
\vskip 12pt%
{\bf Fig.\ \the\figno:} #2\par%
\endinsert\endgroup\par%
\goodbreak%
\global\advance\figno by1%
}
\def\omit#1{}

\def\rem#1{{\sl [#1]}}
\def\pre#1{ ({\tt #1})}
\lref\Pr{J. Propp, {\sl The many faces of alternating-sign matrices}, 
preprint\pre{math.CO/0208125}.}
%
\lref\RS{A.V. Razumov and Yu.G. Stroganov, 
preprint\pre{math.CO/0104216}.}
%
%
\lref\Read{N.\ Read in
  {\sl Proceedings of the Kagom\'e Workshop},
  ed.\ P.\ Chandra (NEC Laboratories, Princeton, 1992).}
\lref\KondevA{J.\ Kondev and C.L.\ Henley,
  {\sl Four-coloring model on the square lattice---a critical ground state},
  Phys.\ Rev.\ B {\bf 52}, 6628--6639 (1995).}
\lref\KondevB{J.\ Kondev and C.L.\ Henley,
  {\sl Kac-Moody symmetries of critical ground states},
  Nucl.\ Phys.\ B {\bf 464}, 540--575 (1996)\pre{cond-mat/9511102}.}
\lref\Batchelor{M.T.\ Batchelor, H.W.J.\ Bl\"ote, B.\ Nienhuis
  and C.M.\ Yang, {\sl Critical behaviour of the fully packed loop model
  on the square lattice},
  J.\ Phys.\ A {\bf 29}, L399--L404 (1996).}
\lref\Raghavan{R.\ Raghavan, C.L.\ Henley and S.L.\ Arouh,
  {\sl New two-color dimer models with critical ground states},
  J.\ Stat.\ Phys.\ {\bf 86}, 517--550 (1997)\pre{cond-mat/9606220}.}
\lref\KondevC{J.\ Kondev,
  {\sl Liouville field theory of fluctuating loops},
  Phys.\ Rev.\ Lett.\ {\bf 78}, 4320--4323 (1997)\pre{cond-mat/9703113}.}
\lref\JacobsenA{J.L.\ Jacobsen and J.\ Kondev,
  {\sl Field theory of compact polymers on the square lattice},
  Nucl.\ Phys.\ B {\bf 532}, 635--688 (1998)\pre{cond-mat/9804048}.}
\lref\KondevD{J.\ Kondev and J.L.\ Jacobsen,
  {\sl Conformational entropy of compact polymers},
  Phys.\ Rev.\ Lett.\ {\bf 81}, 2922--2925 (1998)\pre{cond-mat/9805178}.}
\lref\JacobsenB{J.L.\ Jacobsen and J.\ Kondev,
  {\sl Conformal field theory of the Flory model of protein melting},
  preprint\pre{cond-mat/0209247}.}
\lref\JacobsenC{J.L.\ Jacobsen and J.\ Kondev,
  {\sl Continuous melting of compact polymers},
  preprint\pre{cond-mat/0401504}.}
\lref\Dotsenko{Vl.S.\ Dotsenko, J.L.\ Jacobsen and M.\ Picco,
  {\sl Classification of conformal field theories based on Coulomb gases.
  Application to loop models},
  Nucl.\ Phys.\ B {\bf 618}, 523--550 (2001)\pre{hep-th/0105287}.}
\lref\PZJ{P.~Zinn-Justin, 
  {\sl Non-linear integral equations for complex affine Toda associated to simply laced Lie algebras}, 
  J. Phys. A {\bf 31}, 6747 (1998)\pre{hep-th/9712222}.}
\lref\Hol{T.J.~Hollowood,
{\it Int. J. Mod. Phys.} A8 (1993), 947.}
\lref\Nienhuis{B.\ Nienhuis, {\sl Tiles and colors},
  {\it J.\ Stat.\ Phys.} {\bf 102}, 981--996 (2001)\pre{cond-mat/0005274}.}
\lref\NDC{D.\ Dei Cont and B.\ Nienhuis,
  {\sl The packing of two species of polygons on the square lattice}
\pre{cond-mat/0311244}.}
\lref\Delius{G.W.\ Delius, M.D.\ Gould and Y.-Z.\ Zhang,
  {\sl On the construction of trigonometric solutions of the Yang-Baxter equations},
  {\it Nucl.\ Phys.} B {\bf 432}, 377--403 (1994)\pre{hep-th/9405030}.}
\lref\Resh{N.Y.\ Reshetikhin,
  {\sl A new exactly solvable case of an O($n$) model on a hexagonal lattice},
  J. Phys. A {\bf 24}, 2387--2396 (1991).}
\lref\BazhResh{V.V.~Bazhanov and N.Yu.~Reshetikhin,
{\sl Restricted solid-on-solid models connected with simply laced algebras and 
conformal field theory},
  J. Phys. A {\bf 23}, 1477 (1990).}
\lref\KuResh{P.P.~Kulish and N.Yu.~Reshetikhin,
{\sl Diagonalisation of GL(N) invariant transfer matrices and quantum
$N$-wave system (Lee model)},
  J. Phys. A {\bf 16}, L591 (1983).}
\lref\BVV{O.~Babelon, H.J.~de~Vega and C.-M. Viallet,
 Nucl. Phys. B {\bf 200}, 266 (1982).}
\lref\dV{H.~de~Vega, {\sl Finite-size corrections for nested Bethe Ansatz models
and conformal invariance},
 J. Phys. A {\bf 20}, 6023 (1987);
{\sl Integrable vertex models and extended conformal invariance},
 J. Phys. A {\bf 21}, L1089 (1988).}
\lref\ZP{Y.~Zhou and P.~Pearce, {\sl Solution of functional equations
of restricted $A_{n-1}^{(1)}$ fused lattice models},
 Nucl. Phys. B {\bf 446}, 485--510 (1995).}
\lref\Cardy{H.~W.~J.~Bl\"ote, J.~L.~Cardy and M.~P.~Nightingale,
 {\sl Conformal invariance, the central charge, and universal
 finite-size amplitudes at criticality}, Phys.~Rev.~Lett.~{\bf 56}, 742 (1986);
 I.~Affleck, {\sl Universal term in the free energy at a critical point
 and the conformal anomaly}, Phys.~Rev.~Lett.~{\bf 56}, 746 (1986).}
%
%
\def\FPL{FPL${}^2$}
\def\nb{n_{b}}
\def\nw{n_{w}}
\def\wb{w_{b}}
\def\ww{w_{w}}
\def\boxit#1{\hbox{\vbox{\hrule\hbox{\vrule\vbox to 3pt%
{\vfil\hbox to 3pt{\hfil#1\hfil}\vfil}\vrule%
}\hrule}\hskip-0.4pt}}
\def\young#1{{\vtop{\baselineskip=-1000pt\lineskip=-0.4pt%
\halign{\boxit{##}&&\boxit{##}\crcr%
#1}}\hskip0.4pt}}
\def\fund{{\young{\cr}}}
\def\conj{{\young{\cr\cr\cr}}}
\def\adj{{\young{&\cr\cr\cr}}}
\def\sym{{\young{&\cr}}}
\def\symt{{\young{&\cr&\cr&\cr}}}
\def\anti{{\young{\cr\cr}}}
\def\e#1{{\rm e}^{#1}}
\def\bra#1{\left< #1 \right|}
\def\ket#1{\left| #1 \right>}

\Title{}
{\vbox{
\centerline{Algebraic Bethe Ansatz for the \FPL\ model}
}}\bigskip
\centerline{J. Jacobsen and P. Zinn-Justin}\medskip
\centerline{\it Laboratoire de Physique Th\'eorique et Mod\`eles Statistiques}
\centerline{\it Universit\'e Paris-Sud, B\^atiment 100}
\centerline{\it F-91405 Orsay Cedex, France}
\bigskip

\noindent

An exact solution of the model of fully packed loops of two colors on a square
lattice has recently been proposed by Dei Cont and Nienhuis using the
coordinate Bethe Ansatz approach. We point out here a simpler
alternative, in which the transfer matrix is directly identified as a product
of R-matrices; this allows to apply the (nested) algebraic Bethe Ansatz,
which leads to the same Bethe equations. We comment on some of the applications
of this result.

\Date{02/2004}\def\rem#1{}

\newsec{Introduction}

Models of fluctuating loops play a key role in two-dimensional statistical
physics, and a range of well-known models (Ising, Potts, percolation, O($n$),
to name but a few) can be conveniently studied through their reformulations as
loop models. Exact results about loop models have been produced by a variety
of techniques, including the Coulomb gas, conformal field theory, the Bethe
Ansatz, and stochastic Loewner evolution.

In this note we study the fully packed two-color loop model on the square
lattice (henceforth referred to as the \FPL\ model) from the point of view of
the algebraic Bethe Ansatz. The \FPL\ model was introduced in \KondevA\ as a
generalization of the four-coloring model of the square lattice edges \Read.
It is defined by assigning one of two colors (black or white) to each lattice
edge, subject to the constraint that every vertex be incident to two black and
two white edges. In this way, the black and white edges form fully packed
loops which are given fugacities $\nb$ and $\nw$ depending on their color.

The \FPL\ model has attracted much interest over the last decade. Successive
advances in the Coulomb gas technique have permitted to compute the central
charge and the critical exponents for a number of special cases: the
four-coloring model $(\nb,\nw)=(2,2)$ \refs{\Read,\KondevA}, the dimer-loop
model $(\nb,\nw)=(2,1)$ \Raghavan, and the equal-fugacity case $\nb=\nw$
\KondevC. This eventually led to the solution for general values of $\nb$ and
$\nw$ \KondevD. An interesting special case is that of Hamiltonian walks, with
$(\nb,\nw)=(0,1)$ \JacobsenA. A generalization of the \FPL\ model, obtained by
giving the loops a bending rigidity, was solved in \JacobsenB. It contains as
a special case the so-called Flory model of protein melting \JacobsenC.

All these Coulomb gas results are obtained by making certain reasonable, but
non-rigorous, assumptions about the long wavelength behavior of an associated
interface model. The resulting critical exponents are however believed to be
exact, and they have been successfully tested against numerical Monte Carlo
\refs{\KondevA,\Raghavan} and transfer matrix
\refs{\Batchelor,\JacobsenA,\JacobsenB} results.

To give the results obtained by the Coulomb gas a rigorous status, and to go
beyond it and derive results which are not obtainable from a continuum
approach, it is natural to turn to the methods of integrability. Following
earlier work on the four-coloring model \Nienhuis, Dei Cont and Nienhuis \NDC\
have very recently succeeded in finding a coordinate Bethe Ansatz for the
equal-fugacity \FPL\ model. In particular they have computed the exact
partition function. Moreover, they have shown that when $\nb \neq \nw$, the
\FPL\ model is not integrable, in agreement with earlier expectations
\JacobsenA.

However the Coordinate Bethe Ansatz is a rather complex technique, which,
in order to be made fully rigorous, would require investigation of a large number
of specific configurations.
The goal of the present note is to present a simpler alternative, in which the
transfer matrix of the equal-fugacity \FPL\ model (henceforth we note $n\equiv
\nb=\nw$) is identified with a product of trigonometric R-matrices of
$U_q(\widehat{sl(4)})$ with $n=-q-q^{-1}$ (and an appropriate twist).
Applying the (nested) algebraic Bethe Ansatz allows us to recover the
Bethe equations of Ref.~\NDC\ in a straigthforward fashion. As a bonus we
obtain the central charge and the critical exponents, which are found to agree
with the non-rigorous results of Ref.~\JacobsenA. We also comment on a
$n\to -n$ symmetry of some of the sectors of the transfer matrix.

The paper is organized as follows. The \FPL\ model is defined in Sec.~2.
In Sec.~3 we define its transfer matrix in terms of an R-matrix that adds
four vertices and a twist matrix that takes care of the boundary conditions.
We then show how these matrices are related to those of the affine quantum
group $U_q(\widehat{sl(4)})$ with alternating fundamental and conjugate
representations. The corresponding Bethe Ansatz equations are
discussed in Sec.~4, and we reproduce in particular the eigenvalues of
the transfer matrix \NDC. Finally, in Sec.~5, we give the expressions
of the central charge and conformal weights for $n\le 2$, and we compare
our results to those obtained for fully packed loops on the hexagonal lattice.

\newsec{The \FPL\ model}

Following \JacobsenA, we reformulate the \FPL\ model as a 24-vertex model. For
each vertex of the square lattice, the four incident edges are decorated with
arrows of two possible orientations (outgoing or ingoing with respect to the
vertex) and two possible colors (black or white), subject to the constraint
that each of the four possibilities be represented exactly once around every
vertex. Clearly, following the arrows of a given color traces out an oriented
loop of that color. Note that reversing the arrows along any one loop, and
leaving all other arrows unchanged, leads to another allowed configuration.

\fig\arrows{The six types of vertices in the 24-vertex model (the remaining
vertices are obtained by $\pi/2$ rotations) with their corresponding weights.}
{\epsfxsize=10cm\epsfbox{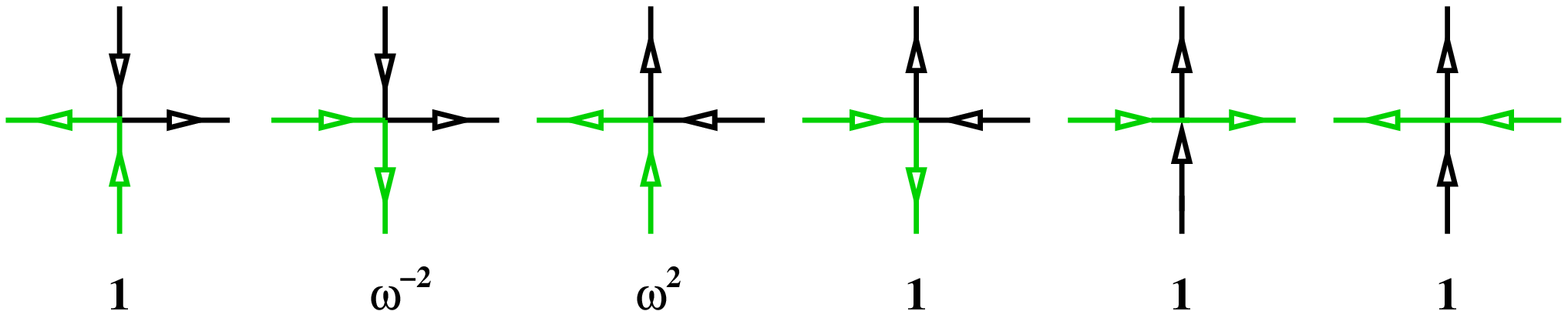}}

We then assign a weight $w=\wb \ww$ to each vertex which is the product of
weights $\wb$ and $\ww$ coming from the oriented black and white loops
respectively. The black weight $\wb=\omega$ (resp.\ $\wb=\omega^{-1}$) when
the black loop makes a right turn (resp.\ a left turn) at the concerned
vertex, and $\wb=1$ when the black loop goes straight. For the white weight
we choose the opposite convention, that is with $\omega$ and $\omega^{-1}$
exchanged.
The weights of the six types of
vertices which are unrelated by $\pi/2$ rotations are given in Fig.~\arrows.

The \FPL\ model with fugacity $n$ for both colors of loops is recovered by
summing independently over the orientations of all loops (black and
white). An anticlockwise (resp.\ clockwise) black loop contributes $\omega^4$
(resp.\ $\omega^{-4}$) to the fugacity, as it must turn four times more
(resp.\ less) to the left than to the right. Thus, $\omega$ is fixed by
\eqn\fugacity{n=\omega^4+\omega^{-4}.}

In order to apply the Bethe Ansatz it is important to specify the boundary
conditions. In the following we shall specialize to the case where the square
lattice is wrapped on a cylinder, i.e., with periodic boundary conditions
across a horizontal row of $2L$ vertices \refs{\JacobsenA,\NDC}. Note that the
argument leading to \fugacity\ only works for contractible loops, i.e., loops
that do not wrap around the periodic direction. To obtain the correct weighing
also for non-contractible loops one introduces a vertical seam separating the
first and the last vertex in each row \NDC. Horizontal edges cutting the seam
are assigned an extra weight of $a$ (resp.\ $a^{-1}$) when covered by a
left-pointing (resp.\ right-pointing) arrow; the convention does {\sl not}
depend on the color of the arrow.
Clearly, non-contractible loops can only
wind once, so $a$ is fixed by
\eqn\winding{n=a+a^{-1}.}

\fig\parity{Parity convention for vertices and edges of the square lattice.
Even (resp.\ odd) edges are shown in dashed (resp.\ solid) linestyle.
Even (resp.\ odd) vertices are shown as dashed (resp.\ solid) circles.}
{\epsfxsize=4cm\epsfbox{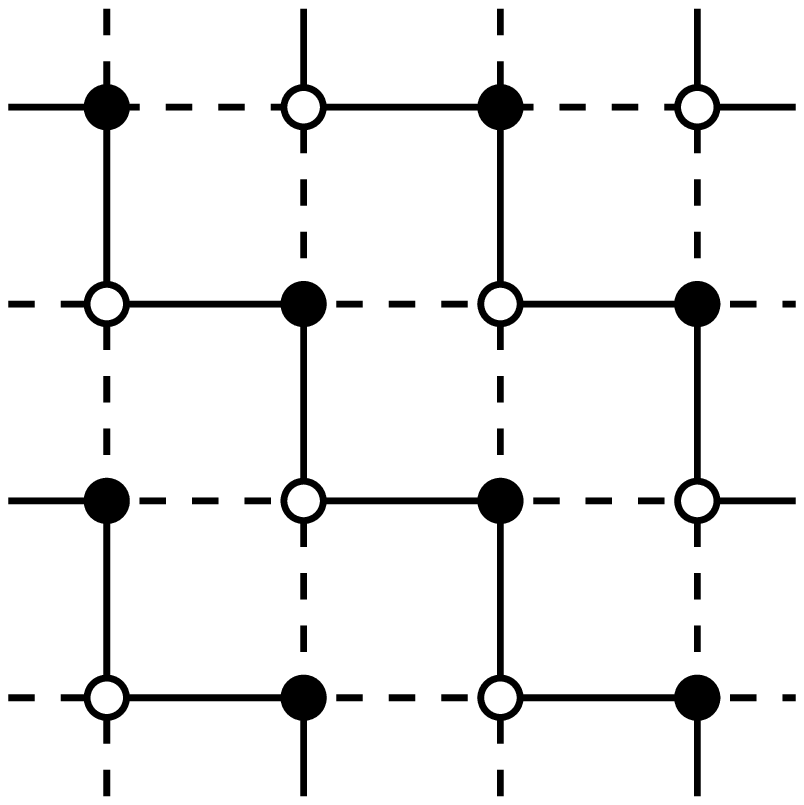}}

With these boundary conditions, the \FPL\ model
contains three conserved quantities \NDC. To explain these, we shall adopt
a convention for the parity of both the vertices and the edges, as shown
in Fig.~\parity. The three components of the conserved charge which are
conserved by the evolution along the cylinder are
\eqn\charges{Q 
=\pmatrix{L \cr L \cr L}-
\pmatrix{N_{w\downarrow}+N_{eb}\cr N_\downarrow\cr N_{w\downarrow}+N_{ob}},}
where $N_{\ldots}$ is the number of vertical edges of a given parity
($e$ = even, $o$ = odd) in the concerned row, $b$ (black) or $w$ (white)
refers to the color of the arrow, and $\uparrow$ (up) or $\downarrow$ (down)
to its orientation. The constant term has been added for convenience.
Strictly speaking, it would make better sense to
talk about conserved charges with respect to a parity convention for the
columns that does not alternate from row to row. In this respect, the charges
\charges\ only commute with the transfer matrix that adds
{\sl two rows} at a time. 

\newsec{The transfer matrix}

\fig\convention{Convention for the labeling of arrow states on an even edge.}
{\epsfxsize=2cm\epsfbox{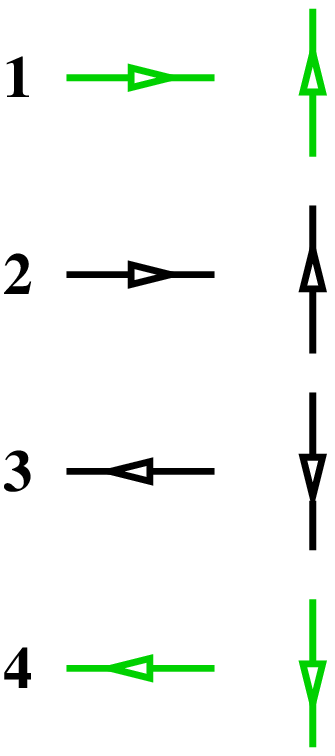}}

Before going on we shall adopt a convention for labeling the arrow state
$i=1,2,3,4$ of each edge in the \FPL\ model. This is shown in Fig.~\convention\
for the case of an even edge; the convention for an odd edge is similar,
but with all arrows reversed (i.e., $i \to 5-i$).

\fig\Tmatrix{The \FPL\ model R-matrix adds two even and two odd vertices as
shown. The arrow indicates the transfer direction. The parity of vertices and
edges follow the conventions of Fig.~\parity. The four dangling edges below
(resp.\ above) the thin dashed line specify the in-state (resp.\ the
out-state).}
{\epsfxsize=4cm\epsfbox{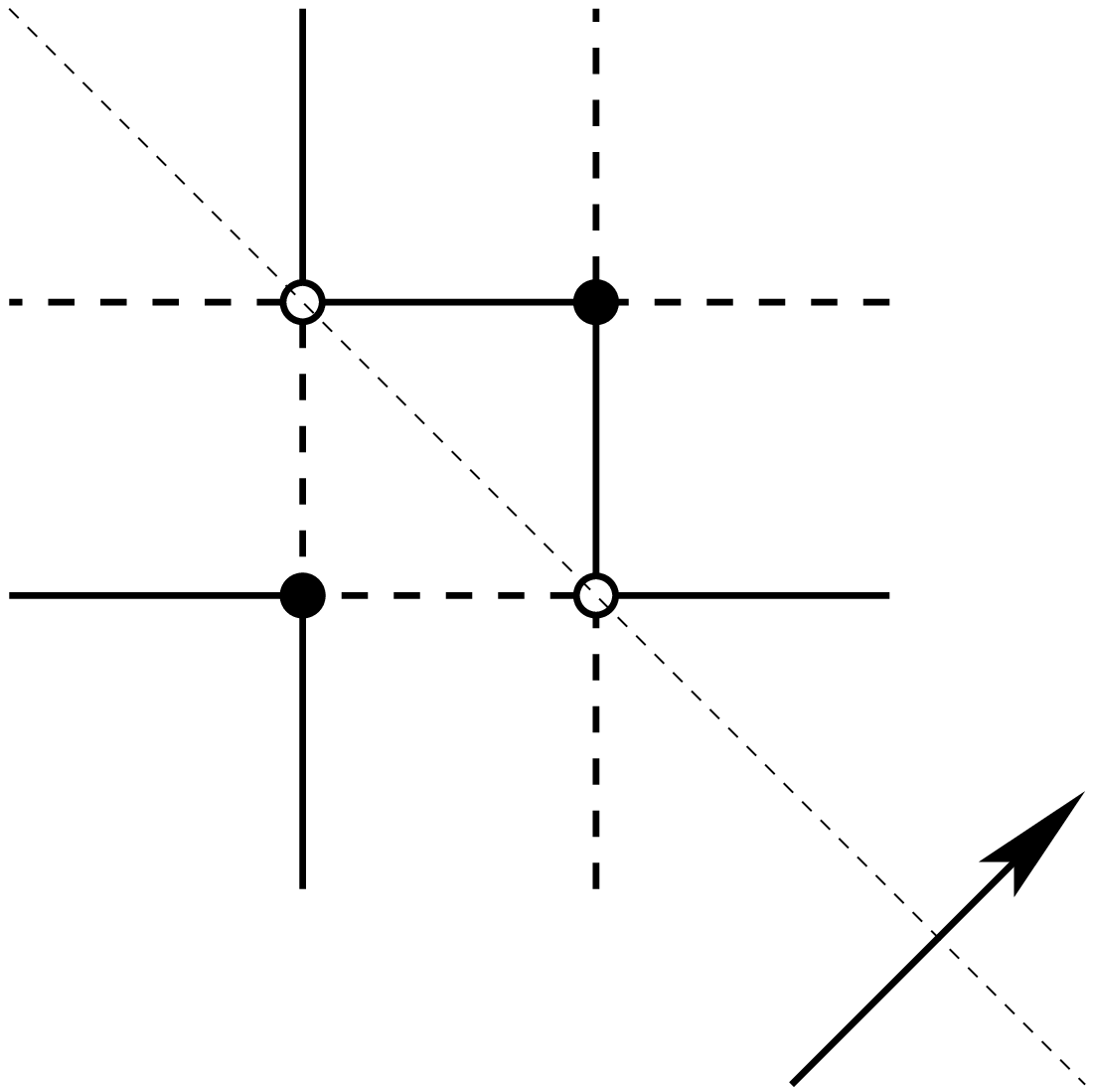}}

We have seen above that from the point of view of the conserved charges, it is
most natural to build up the lattice by adding two rows at a time. We have
also assumed that the horizontal strip is of even width $2L$.
In order to define the row-to-row transfer matrix
$T$, we first define a $256 \times 256$ matrix, which we denote,
in analogy with integrable models, by $R$; it
adds four vertices as shown in Fig.~\Tmatrix. Using the weights of
Fig.~\arrows, it is straightforward to write $R$ explicitly, in the basis
obtained
by using the labeling of Fig.~\convention\ for the external lines and
by taking the tensor product of the corresponding vector spaces.

The transfer matrix that propagates the system in the upwards direction then
reads \eqn\transfer{T = {\rm tr}_a R_{aL}\cdots R_{a2}R_{a1}(\Omega^{-1}\otimes\Omega),} where the
subscript $a$ denotes the ``auxiliary space'' (the pair of horizontal lines)
of dimension 16 and the subscripts $1,2,\ldots,L$ correspond to the $L$ pairs
of vertical lines which form the ``physical space''. The twist $\Omega^{-1} \otimes\Omega$ is
a matrix in the auxiliary space which takes care of the effect of the seam;
explicitly, $\Omega={\rm diag}(1/a,1/a,a,a)$ acts on the upper horizontal line 
whereas $\Omega^{-1}$ acts on the lower line.

\fig\Rmatrix{R-matrix for two rows and two columns, which alternatingly
carry the fundamental representation of $U_q(\widehat{sl(4)})$ and its
conjugate. The arrow indicates the transfer direction.}
{\epsfxsize=4cm\epsfbox{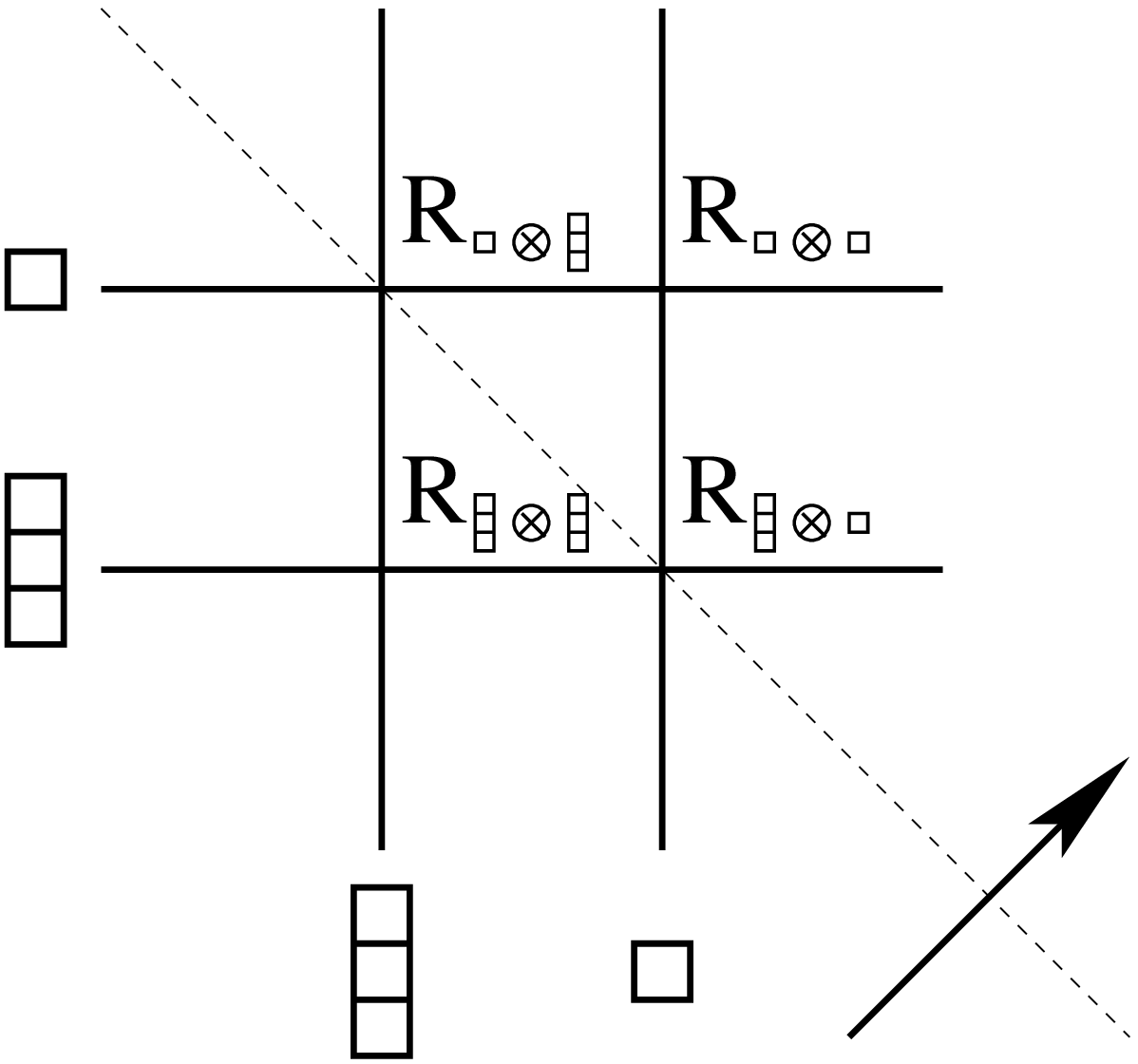}}

We now introduce another R-matrix which is related to the affine quantum
group $U_q(\widehat{sl(4)})$, where $q=-\omega^{-4}$ (so that
$n=-q-q^{-1}$), and which we call ${\bf R}$. It is schematically
described by Fig.~\Rmatrix, in which the two representations
\fund\ and \conj\ of $U_q(\widehat{sl(4)})$ appear
in an alternating fashion.
Conventions are such that at a vertex where two lines intersect,
the first factor in the tensor product refers to the leftmost line of the
in-state, when seen along the transfer direction.

The four R-matrices which appear on Fig.~\Rmatrix\ can be expressed as:
\Delius
\eqna\Rmats
$$\eqalignno{{\bf\check{R}}_{\fund \otimes \fund}(x) &= (qx-q^{-1}x^{-1}){\bf\check{P}}_{\fund\otimes\fund}^\sym + (qx^{-1}-q^{-1}x){\bf\check{P}}_{\fund\otimes\fund}^\anti, &\Rmats a\cr
{\bf\check{R}}_{\fund \otimes \conj}(x) &= (q^2 x-q^{-2}x^{-1})
{\bf\check{P}}_{\fund\otimes\conj}^\adj + (q^2 x^{-1}-q^{-2} x){\bf\check{P}}_{\fund\otimes\conj}^\emptyset, &\Rmats b\cr
{\bf\check{R}}_{\conj \otimes \fund}(x) &= (q^2 x-q^{-2}x^{-1}){\bf\check{P}}_{\conj\otimes\fund}^\adj + (q^2 x^{-1}-q^{-2} x){\bf\check{P}}_{\conj\otimes\fund}^\emptyset, &\Rmats c\cr
{\bf\check{R}}_{\conj \otimes \conj}(x) &= (q x-q^{-1}x^{-1}){\bf\check{P}}_{\conj\otimes\conj}^\symt + (qx^{-1}-q^{-1}x){\bf\check{P}}_{\conj\otimes\conj}^\anti. &\Rmats d\cr
}$$
We have as usual
denoted ${\bf\check{R}}\equiv \Pi {\bf R}$, where $\Pi$ is the operator that permutes the two factors of the tensor product. 
The ${\bf\check{P}}$ are intertwining operators
which can be computed using representation theory;
the parameter $x$ is the ratio of spectral parameters of the horizontal
and vertical line. We shall give the explicit matrix representations
of the R-matrices below, after fixing the values of $x$.

In the present context, one can first define separately transfer matrices
for even and odd rows (however, only their product will be directly related
to the previous transfer matrix $T$). Indeed, we define
${\bf R}_\conj(x)=
{\bf R}_{\conj\otimes\fund}(x/x_{\fund})
{\bf R}_{\conj\otimes\conj}(x/x_{\conj})$
and ${\bf R}_\fund(x)=
{\bf R}_{\fund\otimes\fund}(x/x_{\fund})
{\bf R}_{\fund\otimes\conj}(x/x_{\conj})$,
where the products are meant as in Fig.~\Rmatrix, and the spectral parameters
of vertical lines $x_{\conj}$, $x_{\fund}$ are supposed to be fixed.
Define next
\eqna\twoT
$$
\eqalignno{
{\bf T}_\conj(x)&={\rm tr}_\conj {\bf R}_{\conj L}(x)\cdots {\bf R}_{\conj 2}(x){\bf R}_{\conj 1}(x)
\Omega^{-1}
&\twoT{a}\cr
{\bf T}_\fund(x)&={\rm tr}_\fund {\bf R}_{\fund L}(x)\cdots {\bf R}_{\fund 2}(x){\bf R}_{\fund 1}(x)
\Omega
&\twoT{b}\cr
}
$$
where, as before, the indices determine the spaces on which the matrices
act. Note that $\Omega$ and $\Omega^{-1}$ can be
considered as the same element of the Cartan subalgebra of 
$U_q(\widehat{sl(4)})$, but in fundamental and conjugate representations
respectively.
Due to the Yang--Baxter equation, the ${\bf T}_\conj(x)$ and ${\bf T}_\fund(x)$
form an infinite family of commuting matrices;
their product,
the two-row transfer matrix ${\bf T}(x,y)={\bf T}_\fund(x) {\bf T}_\conj(y)={\bf T}_\conj(y) {\bf T}_\fund(x)$,
is itself of the form
\eqn\oneT{
{\bf T}(x,y)={\rm tr}_a {\bf R}_{aL}(x,y)\cdots {\bf R}_{a2}(x,y)
{\bf R}_{a1}(x,y)
(\Omega^{-1}\otimes \Omega)
}
where ${\bf R}(x,y)={\bf R}_\fund(x) {\bf R}_\conj(y)$.

The claim is that the two R-matrices for two rows and two columns 
$R$ and ${\bf R}$
that we have introduced are 
related. At this point
we choose all horizontal lines (whether odd or even) 
to have the same spectral parameter,
and similarly for all vertical lines, so that the ratio is constant and is:
$x=q^{-1}$. Note that at this special value, the matrices
${\bf\check R}_{\fund\otimes\fund}$
and ${\bf\check R}_{\conj\otimes\conj}$ become 
proportional to projectors onto the
antisymmetric sub-representations 
(this enforces the fact that two loops
of the same color cannot cross each other).

Explicitly: ($c=q^{-1}-q$)
\def\smatrix#1{\null\,\vcenter{%
\baselineskip=7.6pt\mathsurround=0pt\lineskiplimit=0pt\lineskip=0pt%
\ialign{%
\hbox to 16pt{\hfil$\scriptstyle ##$\hfil}&&%
\hbox to 16pt{\hfil$\scriptstyle ##$\hfil}\cr\cr\mathstrut\crcr%
\noalign{\kern-\baselineskip}#1\crcr\mathstrut\crcr\noalign{%
\kern-\baselineskip}}}\,}
\def\psmatrix#1{\left(\smatrix{#1}\right)}
\eqna\Rmat
$$
\eqalignno{
{\bf\check{R}}_{\fund \otimes \fund}(q)&=c\psmatrix{
0 & 0 & 0 & 0 & 0 & 0 & 0 & 0 & 0 & 0 & 0 & 0 & 0 & 0 & 0 & 0 \cr
0 & -q & 0 & 0 & 1 & 0 & 0 & 0 & 0 & 0 & 0 & 0 & 0 & 0 & 0 & 0 \cr
0 & 0 & -q & 0 & 0 & 0 & 0 & 0 & 1 & 0 & 0 & 0 & 0 & 0 & 0 & 0 \cr
0 & 0 & 0 & -q & 0 & 0 & 0 & 0 & 0 & 0 & 0 & 0 & 1 & 0 & 0 & 0 \cr
0 & 1 & 0 & 0 & -q^{-1} & 0 & 0 & 0 & 0 & 0 & 0 & 0 & 0 & 0 & 0 & 0 \cr
0 & 0 & 0 & 0 & 0 & 0 & 0 & 0 & 0 & 0 & 0 & 0 & 0 & 0 & 0 & 0 \cr
0 & 0 & 0 & 0 & 0 & 0 & -q & 0 & 0 & 1 & 0 & 0 & 0 & 0 & 0 & 0 \cr
0 & 0 & 0 & 0 & 0 & 0 & 0 & -q & 0 & 0 & 0 & 0 & 0 & 1 & 0 & 0 \cr
0 & 0 & 1 & 0 & 0 & 0 & 0 & 0 & -q^{-1} & 0 & 0 & 0 & 0 & 0 & 0 & 0 \cr
0 & 0 & 0 & 0 & 0 & 0 & 1 & 0 & 0 & -q^{-1} & 0 & 0 & 0 & 0 & 0 & 0 \cr
0 & 0 & 0 & 0 & 0 & 0 & 0 & 0 & 0 & 0 & 0 & 0 & 0 & 0 & 0 & 0 \cr
0 & 0 & 0 & 0 & 0 & 0 & 0 & 0 & 0 & 0 & 0 & -q & 0 & 0 & 1 & 0 \cr
0 & 0 & 0 & 1 & 0 & 0 & 0 & 0 & 0 & 0 & 0 & 0 & -q^{-1} & 0 & 0 & 0 \cr
0 & 0 & 0 & 0 & 0 & 0 & 0 & 1 & 0 & 0 & 0 & 0 & 0 & -q^{-1} & 0 & 0 \cr
0 & 0 & 0 & 0 & 0 & 0 & 0 & 0 & 0 & 0 & 0 & 1 & 0 & 0 & -q^{-1} & 0 \cr
0 & 0 & 0 & 0 & 0 & 0 & 0 & 0 & 0 & 0 & 0 & 0 & 0 & 0 & 0 & 0 \cr}&\Rmat{a}\cr
{\bf\check{R}}_{\fund \otimes \conj}(q)&=c \psmatrix{
0& 0& 0& 0& 0& -q& 0& 0& 0& 0& -q^3& 0& 0& 0& 0& -q^5\cr 
 0& 0& 0& 0& 1& 0& 0& 0& 0& 0& 0& 0& 0& 0& 0& 0\cr 
 0& 0& 0& 0& 0& 0& 0& 0& 1& 0& 0& 0& 0& 0& 0& 0\cr 
 0& 0& 0& 0& 0& 0& 0& 0& 0& 0& 0& 0& 1& 0& 0& 0\cr 
 0& 1& 0& 0& 0& 0& 0& 0& 0& 0& 0& 0& 0& 0& 0& 0\cr 
 -q^{-1}& 0& 0& 0& 0& 0& 0& 0& 0& 0& -q& 0& 0& 0& 0& -q^3\cr 
 0& 0& 0& 0& 0& 0& 0& 0& 0& 1& 0& 0& 0& 0& 0& 0\cr 
 0& 0& 0& 0& 0& 0& 0& 0& 0& 0& 0& 0& 0& 1& 0& 0\cr 
 0& 0& 1& 0& 0& 0& 0& 0& 0& 0& 0& 0& 0& 0& 0& 0\cr 
 0& 0& 0& 0& 0& 0& 1& 0& 0& 0& 0& 0& 0& 0& 0& 0\cr 
 -q^{-3}& 0& 0& 0& 0& -q^{-1}& 0& 0& 0& 0& 0& 0& 0& 0& 0& -q\cr 
 0& 0& 0& 0& 0& 0& 0& 0& 0& 0& 0& 0& 0& 0& 1& 0\cr 
 0& 0& 0& 1& 0& 0& 0& 0& 0& 0& 0& 0& 0& 0& 0& 0\cr 
 0& 0& 0& 0& 0& 0& 0& 1& 0& 0& 0& 0& 0& 0& 0& 0\cr 
 0& 0& 0& 0& 0& 0& 0& 0& 0& 0& 0& 1& 0& 0& 0& 0\cr 
 -q^{-5}& 0& 0& 0& 0& -q^{-3}& 0& 0& 0& 0& -q^{-1}& 0& 0& 0& 0& 0\cr
}&\Rmat{b}\cr
{\bf\check{R}}_{\conj \otimes \fund}(q)&=c\psmatrix{
0& 0& 0& 0& 0& -q& 0& 0& 0& 0& -q& 0& 0& 0& 0& -q\cr 
 0& 0& 0& 0& 1& 0& 0& 0& 0& 0& 0& 0& 0& 0& 0& 0\cr 
 0& 0& 0& 0& 0& 0& 0& 0& 1& 0& 0& 0& 0& 0& 0& 0\cr 
 0& 0& 0& 0& 0& 0& 0& 0& 0& 0& 0& 0& 1& 0& 0& 0\cr 
 0& 1& 0& 0& 0& 0& 0& 0& 0& 0& 0& 0& 0& 0& 0& 0\cr 
 -q^{-1}& 0& 0& 0& 0& 0& 0& 0& 0& 0& -q& 0& 0& 0& 0& -q\cr 
 0& 0& 0& 0& 0& 0& 0& 0& 0& 1& 0& 0& 0& 0& 0& 0\cr 
 0& 0& 0& 0& 0& 0& 0& 0& 0& 0& 0& 0& 0& 1& 0& 0\cr 
 0& 0& 1& 0& 0& 0& 0& 0& 0& 0& 0& 0& 0& 0& 0& 0\cr 
 0& 0& 0& 0& 0& 0& 1& 0& 0& 0& 0& 0& 0& 0& 0& 0\cr 
 -q^{-1}& 0& 0& 0& 0& -q^{-1}& 0& 0& 0& 0& 0& 0& 0& 0& 0& -q\cr 
 0& 0& 0& 0& 0& 0& 0& 0& 0& 0& 0& 0& 0& 0& 1& 0\cr 
 0& 0& 0& 1& 0& 0& 0& 0& 0& 0& 0& 0& 0& 0& 0& 0\cr 
 0& 0& 0& 0& 0& 0& 0& 1& 0& 0& 0& 0& 0& 0& 0& 0\cr 
 0& 0& 0& 0& 0& 0& 0& 0& 0& 0& 0& 1& 0& 0& 0& 0\cr 
 -q^{-1}& 0& 0& 0& 0& -q^{-1}& 0& 0& 0& 0& -q^{-1}& 0& 0& 0& 0& 0\cr
}&\Rmat{c}\cr
{\bf\check{R}}_{\conj \otimes \conj}(q)&=c\psmatrix{
0 & 0 & 0 & 0 & 0 & 0 & 0 & 0 & 0 & 0 & 0 & 0 & 0 & 0 & 0 & 0 \cr
0 & -q^{-1} & 0 & 0 & 1 & 0 & 0 & 0 & 0 & 0 & 0 & 0 & 0 & 0 & 0 & 0 \cr
0 & 0 & -q^{-1} & 0 & 0 & 0 & 0 & 0 & 1 & 0 & 0 & 0 & 0 & 0 & 0 & 0 \cr
0 & 0 & 0 & -q^{-1} & 0 & 0 & 0 & 0 & 0 & 0 & 0 & 0 & 1 & 0 & 0 & 0 \cr
0 & 1 & 0 & 0 & -q & 0 & 0 & 0 & 0 & 0 & 0 & 0 & 0 & 0 & 0 & 0 \cr
0 & 0 & 0 & 0 & 0 & 0 & 0 & 0 & 0 & 0 & 0 & 0 & 0 & 0 & 0 & 0 \cr
0 & 0 & 0 & 0 & 0 & 0 & -q^{-1} & 0 & 0 & 1 & 0 & 0 & 0 & 0 & 0 & 0 \cr
0 & 0 & 0 & 0 & 0 & 0 & 0 & -q^{-1} & 0 & 0 & 0 & 0 & 0 & 1 & 0 & 0 \cr
0 & 0 & 1 & 0 & 0 & 0 & 0 & 0 & -q & 0 & 0 & 0 & 0 & 0 & 0 & 0 \cr
0 & 0 & 0 & 0 & 0 & 0 & 1 & 0 & 0 & -q & 0 & 0 & 0 & 0 & 0 & 0 \cr
0 & 0 & 0 & 0 & 0 & 0 & 0 & 0 & 0 & 0 & 0 & 0 & 0 & 0 & 0 & 0 \cr
0 & 0 & 0 & 0 & 0 & 0 & 0 & 0 & 0 & 0 & 0 & -q^{-1} & 0 & 0 & 1 & 0 \cr
0 & 0 & 0 & 1 & 0 & 0 & 0 & 0 & 0 & 0 & 0 & 0 & -q & 0 & 0 & 0 \cr
0 & 0 & 0 & 0 & 0 & 0 & 0 & 1 & 0 & 0 & 0 & 0 & 0 & -q & 0 & 0 \cr
0 & 0 & 0 & 0 & 0 & 0 & 0 & 0 & 0 & 0 & 0 & 1 & 0 & 0 & -q & 0 \cr
0 & 0 & 0 & 0 & 0 & 0 & 0 & 0 & 0 & 0 & 0 & 0 & 0 & 0 & 0 & 0 \cr}&\Rmat{d}\cr
}$$
Note that we use the standard bases for $\fund$ and $\conj$ (which are
dual bases of each other).

One can then check that
\eqn\equivR{
{\bf R}=c^4\, U\, R\, U^{-1}
}
where the constant $c^4$ takes cares of the extra factors in Eqs.~\Rmat{}, and
$U$ is a diagonal matrix that
fully factorizes as a tensor product over the four incoming lines:
$U=U_{h\fund}\otimes U_{h\conj}\otimes U_{v\conj}\otimes U_{v\fund}$,
with as a possible choice
\eqn\Us{
\matrix{
U_{h\fund}=\pmatrix{\omega^6&0&0&0\cr 0&\omega^4&0&0\cr 0&0&\omega^2&0\cr 0&0&0&1\cr}
&
U_{h\conj}=\pmatrix{\omega^6&0&0&0\cr 0&\omega^4&0&0\cr 0&0&\omega^2&0\cr 0&0&0&1\cr}
\cr
U_{v\fund}=\pmatrix{\omega^{12}&0&0&0\cr 0&\omega^8&0&0\cr 0&0&\omega^4&0\cr 0&0&0&1\cr}
&
U_{v\conj}=\pmatrix{\omega^{-6}&0&0&0\cr 0&\omega^{-4}&0&0\cr 0&0&\omega^{-2}&0\cr 0&0&0&1\cr}
\cr
}
}

Consequently, the corresponding transfer matrices $T$ and ${\bf T}$ are
also similar up to a constant:
\eqn\equivT{
{\bf T}=c^{4L}\, U_v T U_v^{-1}
}
where $U_v$ is the tensor product of $U_{v\conj}$ and $U_{v\fund}$ for all
vertical lines.

\fig\examples{Examples of nontrivial R-matrix elements in the \FPL\ model.}
{\epsfxsize=8cm\epsfbox{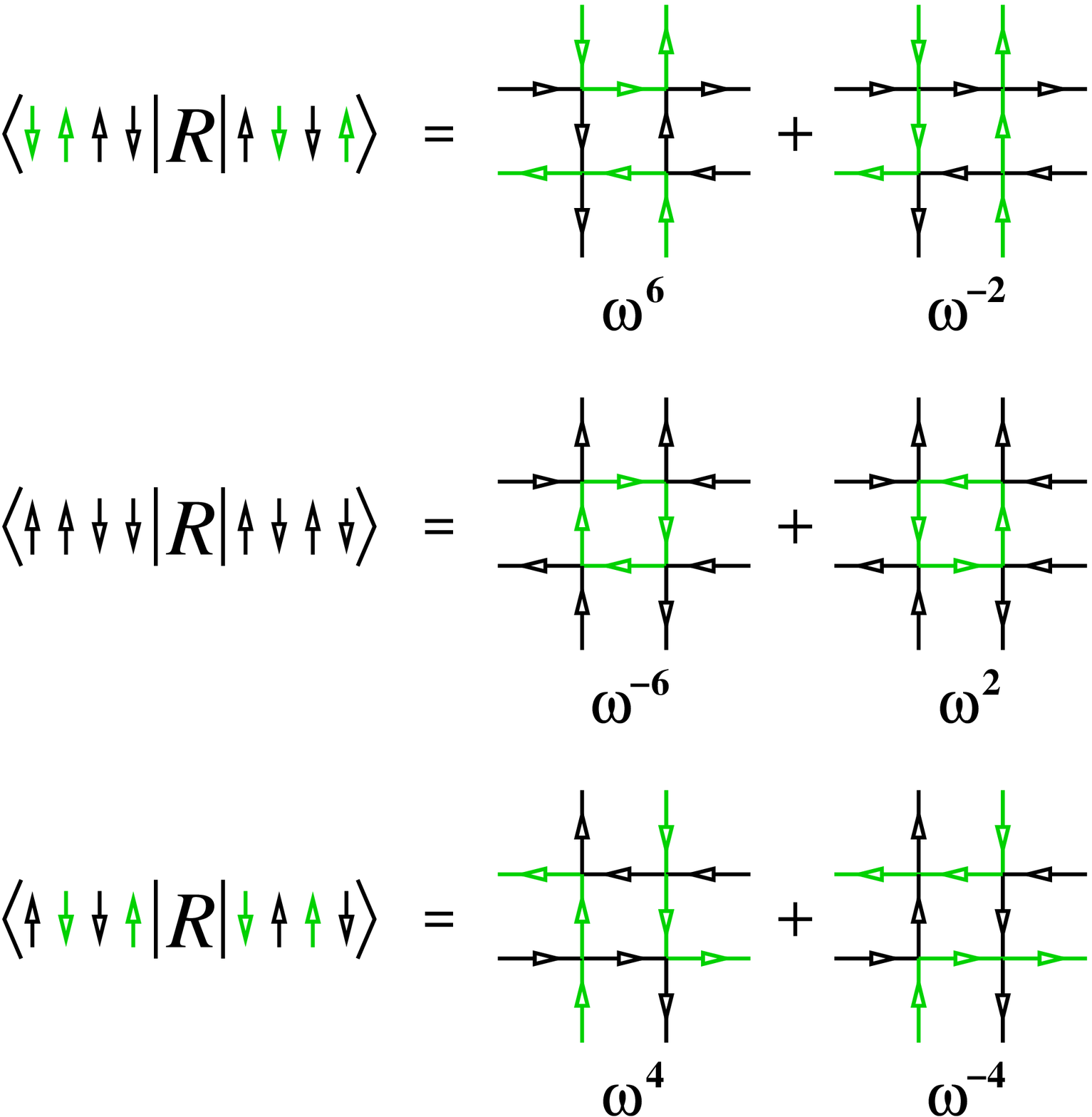}}

Obviously, space does not permit us to reproduce the resulting $256 \times 256$
matrices $R$ and ${\bf R}$. Rather, Fig.~\examples\ gives three examples
of matrix elements. Note that our convention for the R-matrix (see
Fig.~\Tmatrix) is to keep fixed indices for the horizontal and vertical lines.
Thus, if the arrow configuration (coded as in Fig.~\arrows) is
$\rho_4 \rho_3 \rho_2 \rho_1$ for the out-state (read from left to right
when looking along the transfer direction), it is $\rho_2 \rho_1 \rho_4 \rho_3$
for the in-state. 
The three
examples in Fig.~\examples\ then read explicitly:
%
%
$R_{81,18} = \omega^6+\omega^{-2}$;
$R_{103,91} = \omega^{-6}+\omega^2$;
and $R_{239,188} = \omega^4+\omega^{-4}$.
The corresponding entries of ${\bf R}$ are found from \equivR.

\newsec{Algebraic Bethe Ansatz}

The set of commuting transfer matrices ${\bf T}_\fund(x)$, ${\bf T}_\conj(x)$
can be diagonalized using the so-called nested Bethe Ansatz. 
We shall not describe this procedure here and refer to 
\refs{\BVV,\KuResh,\BazhResh} 
for details.
The eigenstates are built by action of operators which depend on parameters
that we call $u^{(i)}_k$, with $i=1,2,3$ and $k=1,\ldots,m^{(i)}$,
on a reference eigenstate (highest weight state) 
which has only white arrows pointing up. 
These parameters satisfy 
equations, which, in our parameterization, are algebraic in the 
$\e{i\gamma u^{(i)}_k}$, where $\gamma$ is such that $q=-\e{-i\gamma}$.
Explicitly, call $\omega^{(i)}$ 
the diagonal elements of the twist $\Omega$ in the
fundamental representation (here, $\omega^{(1)}=\omega^{(2)}=1/a$, $\omega^{(3)}=\omega^{(4)}=a$), with $\prod_{i=1}^4 \omega^{(i)}=1$;
and define 
the functions $Q^{(i)}(u)=\prod_{k=1}^{m^{(i)}} {\sin\gamma(u-u^{(i)}_k)}$,
$i=1,2,3$. $Q^{(0)}\equiv Q^{(4)}\equiv 1$.
Then the Bethe Ansatz equations read
\eqn\BAE{
-{Q^{(i+1)}(u_k^{(i)}+1)\over Q^{(i+1)}(u_k^{(i)})}
{Q^{(i)}(u_k^{(i)}-1)\over Q^{(i)}(u_k^{(i)}+1)}
{Q^{(i-1)}(u_k^{(i)})\over Q^{(i-1)}(u_k^{(i)}-1)}
={\omega^{(i)}\over\omega^{(i+1)}} {f^{(i)}(u_k^{(i)})\over f^{(i+1)}(u_k^{(i)})}
}
for $1\le i\le 3$, $1\le k\le m^{(i)}$.
The functions $f^{(i)}$ depend on the representations and spectral parameters
of the physical space (and on the twist); here, one easily computes 
$f^{(1)}(u)=\omega^{(1)}(\sin \gamma u \sin\gamma(u-1))^L $, 
$f^{(i)}(u)=\omega^{(i)}(\sin \gamma(u+1)\sin\gamma(u-1))^L$ for $i=2,3$, 
$f^{(4)}(u)=\omega^{(4)}(\sin \gamma(u+1)\sin\gamma u)^L$.

The corresponding eigenvalues of ${\bf T}_\conj(u)$ and ${\bf T}_\fund(u)$, 
in the parameterization $x=-\e{i\gamma (u+1)}$,
are
\eqna\eig
$$\eqalignno{
t_\fund(u)&=X^{(1)}(u)+X^{(2)}(u)+X^{(3)}(u)+X^{(4)}(u)&\eig{a}\cr
t_\conj(u)&=\tilde{X}^{(1)}(u)+\tilde{X}^{(2)}(u)+\tilde{X}^{(3)}(u)+\tilde{X}^{(4)}(u)
&\eig{b}\cr
}$$
where $X^{(i)}(u)={Q^{(i-1)}(u-1)\over Q^{(i-1)}(u)} 
{Q^{(i)}(u+1)\over Q^{(i)}(u)} f^{(i)}(u)$ 
and $\tilde{X}^{(i)}(u)={Q^{(i-1)}(u+i-2)\over Q^{(i-1)}(u+i-3)}
{Q^{(i)}(u+i-3)\over Q^{(i)}(u+i-2)} f^{(5-i)}(u)$.

We now choose $u=0$, so that
the $X^{(1)}$, $X^{(4)}$, $\tilde{X}^{(1)}$, $\tilde{X}^{(4)}$ 
vanish (this, once again, can be interpreted as 
a consequence of the requirement that
two loops of the same color do not cross each other).
Finally, we consider the two row-matrix ${\bf T}={\bf T}_\fund {\bf T}_\conj$. 
Its eigenvalue is obtained by
taking the product of the remaining terms in Eqs.~\eig{}; 
rewriting the $X^{(i)}$ as functions
of $Q^{(i)}$, getting rid of the extra factors $\sin\gamma$ which compensate
the factors of $c$ in Eq.~\equivT, we find the eigenvalues of $T$ to be
\eqnn\eigfinal
$$\eqalignno{
t &= 2 {Q^{(2)}(1)Q^{(2)}(-1)\over Q^{(2)}(0)^2}
 + {\omega^{(2)}\over\omega^{(3)}}
\left( {Q^{(2)}(1)\over Q^{(2)}(0)} \right)^2 
   {Q^{(1)}(-1)\over Q^{(1)}(0)} {Q^{(3)}(0)\over Q^{(3)}(1)}
 + {\omega^{(3)}\over\omega^{(2)}}
\left( {Q^{(2)}(-1)\over Q^{(2)}(0)}\right)^2 
   {Q^{(1)}(0)\over Q^{(1)}(-1)} {Q^{(3)}(1)\over Q^{(3)}(0)}\cr
 &= \left(a^{-1}
 {Q^{(2)}(1)\over Q^{(2)}(0)} 
\sqrt{   {Q^{(1)}(-1)\over Q^{(1)}(0)} {Q^{(3)}(0)\over Q^{(3)}(1)}}
+ a\,
{Q^{(2)}(-1)\over Q^{(2)}(0)} 
\sqrt{   {Q^{(1)}(0)\over Q^{(1)}(-1)} {Q^{(3)}(1)\over Q^{(3)}(0)}}
 \right)^2&\eigfinal\cr
}
$$
where in the last line we have used the explicit expression of the twist.
This is precisely the square of the expression found in \NDC\ for the one-row
transfer matrix. The correspondence of notations is as follows:
\eqn\nota{
u_k\equiv 2i (u^{(1)}_k+1/2)\qquad
v_k\equiv 2i (u^{(3)}_k-1/2)\qquad
w_k\equiv 2i u^{(2)}_k
}

Furthermore, the Cartan subalgebra produces three conserved quantities; 
in fundamental and conjugate representations, they have the following
expression: (basis of the dual of the root lattice)
\eqn\BAcharge{
{Q_{1\fund}\atop -Q_{1\conj}}\Big\}=\pmatrix{1&0&0&0\cr 0&0&0&0\cr 0&0&0&0\cr 0&0&0&0}
-{\textstyle{1\over4}} I
\ 
{Q_{2\fund}\atop -Q_{2\conj}}\Big\}=\pmatrix{1&0&0&0\cr 0&1&0&0\cr 0&0&0&0\cr 0&0&0&0}
-{\textstyle{1\over2}} I
\ 
{Q_{3\fund}\atop -Q_{3\conj}}\Big\}=\pmatrix{1&0&0&0\cr 0&1&0&0\cr 0&0&1&0\cr 0&0&0&0}
-{\textstyle{3\over 4}} I
}
Combining this with the correspondence given by Fig.~3, it is easy
to identify them with the components of the charge $Q$ of Eq.~\charges.
Their value is determined by
the numbers $m^{(i)}$ of Bethe roots of kind $i$: 
each root of the kind $i$
decreases by one the $i^{\rm th}$ component of the charge,
starting from the reference state which has $Q=\pmatrix{L\cr L\cr L}$. Comparing
with Eq.~\charges, we deduce that

\eqn\BAchargeb{
m^{(1)}=N_{w\downarrow}+N_{eb}
\qquad
m^{(2)}=N_\downarrow
\qquad
m^{(3)}=N_{w\downarrow}+N_{ob}
}

Note that if $m^{(1)}>0$ but $m^{(3)}=0$, only even arrows are modified compared
to the reference state (i.e.\ all odd arrows are white pointing up); and similarly
for $m^{(3)}>0$, $m^{(1)}=0$ and odd arrows.

\newsec{Results and conclusions}
We have found that the nested algebraic Bethe Ansatz can
be used to solve the \FPL\ model at $\nb=\nw$. The latter is therefore identified
with a standard integrable vertex model associated to $U_q(\widehat{sl(4)})$,
and its transfer matrix embedded into an infinite set of commuting transfer 
matrices; and many results follow immediately.

In particular, the long distance behavior of this type of models has been
studied by many methods (see for instance \refs{\dV,\ZP,\PZJ}). For $|n| > 2$
the spectrum has a gap and the correlation length is finite. In the following
we focus instead on the critical regime $|n| \le 2$, and we parameterize
$n=2\cos\gamma$.

As our model is isotropic, we expect the largest eigenvalue
of the transfer matrix to have the following asymptotic behavior \Cardy
\eqn\fundasy{
\log t_0(L)= - L f_0 + {\pi c\over 6L} + \cdots\qquad {\rm for \ }L\to\infty,
}
where $c$ is the central charge of the infra-red conformal field theory.
Here note that only fundamental representations (\fund\ and
\conj) are used, so we are dealing with a non-fused model. 
Assuming the usual form for the ground state,
standard computations (see e.g.\ \PZJ) lead, in this type of models,
to the following form of $c$
\eqn\cc{
c=r-{3\over \pi(\pi-\gamma)} \bra{w} C^{-1} \ket{w}
}
where $r$ and $C$ are respectively the rank and the Cartan matrix of the
underlying algebra, and $w$ is a vector with components
$w_s={1\over i}\log(\omega^{(s)}/\omega^{(s+1)})$ that
parameterizes the twist. If we now specialize to $A_3$ and to our choice of
boundary conditions: $w_1=w_3=0$, $w_2=-2\gamma$, we obtain
\eqn\ccb{
c=3-12{\gamma^2\over\pi(\pi-\gamma)},
}
which coincides with what was found in \KondevC. Note that this is not the 
central charge of the $W(A_3)$ conformal field theory---the latter
can also be obtained within the framework of this model, but with a different twist.

One can also investigate the nature of excitations above the ground state.
They are, of course, gapless and describe solitons associated to the three 
fundamental representations of $A_3$ interacting with the standard S-matrices
\refs{\PZJ,\Hol},
plus possible bound states in certain regimes of $\gamma$.
In the infra-red limit the dispersion relation can be linearized and the
corresponding low-lying excited states are related to 
the conformal weights $\Delta_n$ of the aforementioned CFT via
\eqn\excasy{
\log t_n(L)= - L f_0 + {\pi (c-24\Delta_n)\over 6L} + \cdots\qquad {\rm for \ }L\to\infty.
}
One can check that the weights thus obtained fit with the formulae
of the Coulomb gas picture:
\eqn\confwei{
\Delta_n= {1\over 4}\bra{e}K^{-1}\ket{e-2e_0}+{1\over 4} \bra{m} K \ket{m}
}
where $K={1\over2}(1-\gamma/\pi) C$, $e$ (resp.\ $m$) is the
electric (resp.\ magnetic) charge which belongs to the lattices of weights
(resp.\ roots) of $A_3$, and $e_0$ is the background charge, related to our twist
by $e_0={1\over 2\pi} w$. 
This constitutes a confirmation of the results of \KondevD.
Incidentally, the fact that for $\nb\ne \nw$ the quadratic form $K$ appearing
in the conformal weights as given in \KondevD\ 
is generically not related to a Cartan matrix can
be considered an indication of the non-integrability of the model.

We have made some numerical checks of the structure of the ground state and
excited state described above. As it turned out, this picture was confirmed
for $n\ge 0$; however, for $n<0$ the ground state of the usual form (with,
using the notations of Eq.~\nota, real $u_k$, $v_k$, $w_k$) is {\it not}\/ the
state corresponding to the dominant eigenvalue of the transfer matrix. Indeed,
we have found a $n\to -n$ symmetry of the eigenvalue spectrum corresponding to
the sectors where both $L-m^{(2)}$ and $m^{(1)}+m^{(3)}$ are even,
cf.~Eq.~\BAchargeb; this applies thus in particular to the ground state sector
which has $m^{(1)}=m^{(2)}=m^{(3)}=L$. This symmetry can be described in the
Bethe Ansatz equations as the transformation of the Bethe roots (with the
notations of Eq.~\nota):
\eqn\ntominusn{
\gamma u_k \to -(\pi-\gamma) u_k\qquad \gamma v_k\to -(\pi-\gamma) v_k\qquad
\gamma w_k\to -(\pi-\gamma) w_k+i\pi,
}
where we recall that $n=2\cos\gamma$, so that $-n=2\cos(\pi-\gamma)$.
One can check that this transformation
leaves the eigenvalues \eigfinal\ invariant. In particular, we
conclude that the ``real'' ground state eigenvalue $t_0$ only depends on $|n|$, and identifies
with the one described above only for $n \ge 0$. For $n<0$, 
the ``fake'' ground state eigenvalue corresponds
to a state very high above the real ground state (even the bulk part 
being different as $L\to\infty$),
whose only special property is that it is the analytic continuation of the $n>0$ ground state
eigenvalue. 
At the moment, we do not have a satisfactory explanation of this phenomenon. The non-unitarity
of the $n<0$ theory, or our boundary conditions (we can only consider the theory on
a cylinder, but not on a torus
due to the issue of winding loops) might play some role in it.

[We remark parenthetically that the symmetry of the ground state sector holds
true more generally for the $n_b \neq n_w$ \FPL\ model, under the
transformation $(n_b,n_w) \to (-n_b,-n_w)$. This is based on the observation
that, with suitable periodic boundary conditions, all terms in the
high-fugacity expansion of the partition function have $N_b+N_w$ even, where
$N_b$ (resp.\ $N_w$) is the number of black (resp.\ white) loops. To see this,
represent the dominant state at $n_b,n_w \to \infty$ as an ``ideal state'' in
the four-coloring picture \KondevA, with black (resp.\ white) loops being an
alternation of colors 1 and 2 (resp.\ 3 and 4). Note that the high-fugacity
expansion of the 1--2 (black) and 3--4 (white) loops (disregarding their
orientation) can be obtained by only permuting the colors around the two {\sl
other} types of small loops (say, of types 1--3 and 2--4), and that these loops
stay of length four. Examining all possible 1--2 and 3--4 loop environments of
a plaquette occupied by the 1--3 and 2--4 loops proves our statement.]

It is interesting to compare our results with the work of Reshetikhin
\Resh\ on fully-packed loops on the hexagonal lattice. After contracting
the vertices of the hexagonal lattice two by two, so as to form a square
lattice, this author identified the R-matrix with that of the integrable
model $U_q(\widehat{sl(3)})$. The choice of spectral parameter, as in our case,
makes ${\bf\check{R}}$ degenerate into a projection operator.
However, there are 
important differences.
First, in \Resh\ the underlying algebra is of course different and
all horizontal and vertical lines carry the same
representation \fund\ of $U_q(\widehat{sl(3)})$, in contrast to
the alternation of \fund\ and \conj\ used here. Second, in \Resh\ there
is no twist $\Omega$ in the auxiliary space, whence contractible and
non-contractible loops carry respective weights of $n$ and $2$. In
particular, for $n\le 2$, the central charge is constant, $c=2$.
When $n=2$, the continuum limit of the hexagonal-lattice loop model becomes a
$SU(3)_{k=1}$ free field Wess-Zumino-Witten theory; this is a consequence of
the $U_q(\widehat{sl(3)})$ identification and of the fact that only the
fundamental representation is used. Likewise, the $U_q(\widehat{sl(4)})$
identification of the \FPL\ model reported in the present work implies that
the $n=2$ case is a $SU(4)_{k=1}$ WZW theory in the continuum limit. Indeed,
the four-coloring model was originally constructed by Read \Read\ so as to
have a $SU(4)_{k=1}$ symmetry.

%
%
%

\listrefs
\bye